\documentclass[fleqn,usenatbib]{mnras}

\usepackage{newtxtext,newtxmath}
\usepackage[T1]{fontenc}

\DeclareRobustCommand{\VAN}[3]{#2}
\let\VANthebibliography\thebibliography
\def\thebibliography{\DeclareRobustCommand{\VAN}[3]{##3}\VANthebibliography}

\usepackage{graphicx}	% Including figure files
\usepackage{amsmath}	% Advanced maths commands

\usepackage{xcolor}

\newcommand*{\msun}{\ensuremath{\rm{M}_{\odot}}\text{ }}

\newcommand*    \pc{{\,\mathrm{pc}}}
\newcommand*    \kpc{{\,\mathrm{kpc}}}

\newcommand*    \gyr{{\,\rm Gyr}}
\newcommand*    \mbh{M_{\rm bh}}

\title[Eccentricity evolution of PTA sources]{Eccentricity evolution of PTA sources from cosmological initial conditions} %{MNRAS \LaTeXe\ template -- 

\author[F. Fastidio et al.]{
F. Fastidio,$^{1,2}$\thanks{E-mail: f.fastidio@campus.unimib.it}
A. Gualandris,$^{2}$
A. Sesana$^{1,3,4}$
E. Bortolas$^{1,3,4}$
and W. Dehnen$^{5}$
\\
$^{1}$Dipartimento di Fisica ``G. Occhialini", Universit{\'a} degli Studi di Milano-Bicocca, Piazza della Scienza 3, I-20126 Milano, Italy\label{unimib}\\
$^{2}$School of Mathematics and Physics, Faculty of Engineering and Physical Science, University 
    of Surrey, Guildford GU2 7XH, UK\label{surrey}\\
$^{3}$INAF - Osservatorio Astronomico di Brera, via Brera 20, I-20121 Milano, Italy\label{inaf-brera}\\
$^{4}$INFN, Sezione di Milano-Bicocca, Piazza della Scienza 3, I-20126 Milano, Italy\label{infn-unimib}\\
$^{5}$Astronomisches Rechen-Institut, Zentrum f{\"u}r Astronomie der Universit{\"a}t Heidelberg, M{\"o}nchhofstra\ss{}e 12-14, 69120, Heidelberg, Germany
}

\date{Accepted XXX. Received YYY; in original form ZZZ}

\pubyear{2024}

\begin{document}
\label{firstpage}
\pagerange{\pageref{firstpage}--\pageref{lastpage}}
\maketitle

\begin{abstract}
Recent results from pulsar timing arrays (PTAs) show evidence for a gravitational wave background (GWB) consistent with a population of unresolved supermassive black hole (SMBH) binaries (BHBs). While the data do not yet constrain the slope of the spectrum, this appears to flatten at the lowest frequencies, deviating from the power-law shape expected for circular binaries evolving solely due to gravitational wave (GW) emission. 
Interestingly, such flattening can be explained with a population of eccentric rather than circular binaries. The eccentricity of BHBs is notoriously difficult to predict based simply on the parameters of the host galaxies and the initial galactic orbit, as it is subject to stochastic effects. We study the evolution of the eccentricity of BHBs formed in galactic mergers with cosmological initial conditions from pairing to coalescence, with a focus on potential PTA sources. We select galactic mergers from the IllustrisTNG100-1 simulation and re-simulate them at high resolution with the $N$-body code {\tt Griffin} down to binary separations of the order of a parsec. We then estimate coalescence timescales with a semi-analytical model of the evolution under the effects of GW emission and stellar hardening. We find that most mergers in IllustrisTNG100-1 occur on highly eccentric orbits, and that the eccentricity of BHBs at binary formation correlates with the initial eccentricity of the merger, if this is no larger than approximately 0.9. For extremely eccentric mergers, the binaries tend to form with modest eccentricities. We discuss the implications of these results on the interpretation of the observed GWB. 
\end{abstract}


%%%%%%%%%%%%%%%%%%%%%%%%%%%%%%%%%%%%%%%%%%%%%%%%%%

%%%%%%%%%%%%%%%%% BODY OF PAPER %%%%%%%%%%%%%%%%%%

\section{Introduction}

Supermassive black holes (SMBHs) are expected to ubiquitously reside at the centre of massive galaxies \citep[e.g.][]{1995ARA&A..33..581K, 2013ARA&A..51..511K}. In the $\Lambda$CDM scenario, hierarchical structure formation implies that galaxies grow through consecutive mergers \citep{1977ApJ...217L.125O, 1980MNRAS.191P...1W, 1993MNRAS.262..627L}; these encounters naturally lead to the formation of SMBH binaries (BHBs) \citep{1980Natur.287..307B}. 
When the two progenitor galaxies merge, their respective SMBHs form an initially unbound pair \citep{1992ARA&A..30..705B}. The two SMBHs then undergo a three-phase evolution, where different mechanisms extract energy and angular momentum from the system \citep{1980Natur.287..307B}.

The first phase is dynamical friction against the dark matter (DM) and stars \citep{1943ApJ....97..255C}, which is typically efficient from tens of kpc down to pc-scale separations. When the mass enclosed within the orbit of the pair becomes comparable to its total mass, the two SMBHs become gravitationally bound \citep{2005LRR.....8....8M} and dynamical friction becomes less efficient. The second phase relies on encounters with individual stars, which harden the binary through the gravitational slingshot effect \citep{1996NewA....1...35Q, 2006ApJ...651..392S}. This process can lead to two different outcomes depending on the supply of stars on low angular momentum orbits that can interact with the binary. If the supply of stars is insufficient to continue hardening the binary, the BHB stalls and fails to merge within a Hubble time \citep{1980Natur.287..307B}, giving rise to the "Final-Parsec Problem." If the scattering process is efficient, the binary continues to harden past the parsec scale until gravitational wave (GW) emission takes over. It has been demonstrated that binary stalling is not likely to happen in nature: simulations show that merger remnants that host BHBs are usually triaxial \citep{2016ApJ...828...73K, 2018MNRAS.477.2310B} and this triggers collisionless refilling of the loss cone (i.e. the region in phase space leading to interaction with the binary) on a time scale shorter than the Hubble time \citep{2006ApJ...642L..21B, 2015ApJ...810...49V, 2017MNRAS.464.2301G}. Because the total angular momentum of stars is not conserved in such potentials, torques lead to angular momentum diffusion and a replenishment of the centrophilic orbits previously depleted by slingshot ejections \citep{2002MNRAS.331..935Y}. Still, depending on the properties of the galaxy, the hardening process may take longer than a Gyr.
%Moreover, we have just a few observational candidates of BHBs \citep[e.g.][]{2012AdAst2012E...3D}, whilst we do have evidence of single SMBHs at the centre of most galaxies, which seems to validate the scenario in which binaries are efficiently brought to coalescence. 
%\noindent Finally, the 
The third binary-evolution phase is driven by GW emission. If loss cone refilling is efficient and the SMBHs reach separations of order a mpc, GW emission quickly drives the binary to coalescence \citep{1964PhRv..136.1224P}. The superposition of the GW signals coming from many unresolved BHBs produces a gravitational wave background \citep[GWB;][]{1995ApJ...446..543R, 2003ApJ...583..616J,2008MNRAS.390..192S}, that can be observed by pulsar timing arrays \citep[PTAs;][]{1990ApJ...361..300F} by detecting correlated deviations in the time of arrivals (TOAs) of radio signals from an ensamble of millisecond pulsars. 

The shape of the GWB is strongly dependent on the processes that bring the black holes to coalescence and on the binary's orbital eccentricity: whilst a population of circular binaries evolving solely due to GW emission would produce, on average, a power-law spectrum of the type $S(f)\propto f^{-13/3}$ \citep{2001astro.ph..8028P, 2015MNRAS.453.2576L},  environmental coupling (i.e. interactions with stars and gas) will cause a flattening or even a turnover at low frequencies \citep{2011MNRAS.411.1467K,2014MNRAS.442...56R}. Moreover, if we account for eccentric binaries, two other effects will be seen on the GWB: (i) emission at higher frequencies will be boosted due to the emission at harmonics higher than twice the orbital frequency; (ii) the close pericentric passages resulting from large eccentricities will lead to a very fast inspiral and therefore to an overall attenuation of the emission at all frequencies (in addition to a shorter merger timescale) \citep[e.g.][]{2017MNRAS.470.1738C, 2017MNRAS.471.4508K}. The combination of the two effects, together with the fact that GW emission tends to circularise the orbit as the pair shrinks, results in a suppression of the low frequency portion of the spectrum with only a slight increase at high frequencies \citep{2015ASSP...40..147S}. 

Recent results from all PTA collaborations (i.e. European PTA, Indian PTA, Parkes PTA, North America Nanohertz Observatory for GWs (NANOGrav) and Chinese PTA)  have shown evidence for a GWB, with statistical significance between 2-4 $\sigma$ \citep{2023A&A...678A..48E,2023A&A...678A..49E,2023A&A...678A..50E,2024A&A...685A..94E,2023arXiv230616226A,2023PhRvL.131q1001S,2023ApJ...951L...9A,2023ApJ...951L..10A,2023ApJ...951L..50A,2023ApJ...951L...8A,2023ApJ...951L..11A,2023ApJ...951L...6R,2023RAA....23g5024X}.
When modelled as a single power law, the slope of the observed GWB appears to 
be slightly flatter than the $S\propto f^{-13/3}$ expected in the circular, GW driven case. However, the detection significance is still low and it is hard to accurately determine the properties of the signal \citep[see e.g.][]{2024A&A...685A..94E}. While data are currently not sufficiently constraining to draw definitive conclusions on the properties of the BHB population that could produce the observed signal, these first results highlight the importance of accurately studying the eccentricity evolution of BHBs down to the GW inspiral phase. 

The dynamical evolution of BHBs from pairing to coalescence and its dependence on eccentricity have been studied by means of $N$-body simulations \citep[e.g.][]{2015ApJ...810...49V, 2016MNRAS.461.1023B, 2017MNRAS.464.2301G, 2022MNRAS.511.4753G}. \citet{2022MNRAS.511.4753G} find that the eccentricity at binary formation, though affected by stochasticity due to encounters with stars, preserves a strong correlation with the initial orbital eccentricity of the galactic merger. More specifically, the eccentricity of the unbound SMBH pair tends to decrease during the dynamical friction phase, while during binary hardening it increases in minor mergers (unless the binary is already approximately circular) and remains almost unchanged in major mergers. In addition, \citet{2020MNRAS.497..739N} show that the scatter in eccentricity at binary formation can be significantly reduced by increasing the mass resolution of the simulations, as the scatter is an artefact of poor phase-space discretisation. However,  \citet{2023MNRAS.526.2688R} argue that the scatter in eccentricity is a consequence of the sensitivity to perturbations of the nearly radial trajectories that SMBHs travel along right before binary formation: they find that, for initial orbital eccentricities $e_0=0.99$, binary eccentricities at formation can span the whole range $[0,1]$, with a weak dependence on mass resolution. 

This paper has two main goals: (i) determine the typical initial eccentricity (i.e. final eccentricity of galactic mergers) of PTA-like sources by performing a statistical study on merger trees drawn from the cosmological simulation IllustrisTNG100-1 \citep{2018MNRAS.475..676S, 2018MNRAS.475..648P, 2018MNRAS.475..624N, 2018MNRAS.477.1206N, 2018MNRAS.480.5113M}; (ii) study the eccentricity evolution of a selected sub-sample of these mergers, adopting initial conditions from IllustrisTNG100-1 and following the dynamics through dynamical friction, hardening and finally GW emission phase.

Since cosmological simulations have intrinsically low resolution, we extract merger and galactic properties at early times, when the SMBHs are at separations of tens of kpc, and generate high resolution realisations of the interacting galaxies. We then model the galactic merger with the state-of-the-art Fast Multiple Method (FMM) code {\tt Griffin} \citep{2014ComAC...1....1D} and follow the dynamics of the BHB to pc-scale separations. Finally, we extrapolate the evolution down to the GW emission phase and coalescence via a semi-analytical model \citep[SAM;][]{2006ApJ...651..392S, 2010ApJ...719..851S}. While the semi-major axis of the binary shrinks during both the hardening phase and the GW-driven decay, the binary eccentricity grows only in the former stage (more so if the initial eccentricity is high) and the binary circularises quickly once GW emission becomes efficient. Correctly predicting the eccentricity at the end of the hardening phase is thus of primary importance, since it determines the onset and the duration of the GW-driven decay. 

We find that the majority of galactic mergers happen on almost radial orbits ($e_0\sim0.97$) and the BHBs that form in the merger remnant appear to have a different evolution and merger timescale depending on the initial eccentricity $e_0$ of the galactic merger: (i) if $e_0\lesssim0.9$ BHBs form with highly eccentric orbits; (ii) if $e_0\gtrsim 0.9$ BHBs tend to settle on more circular orbits.

The paper is structured as follows. In Section \ref{sec:methods} we describe the methods, reporting (i) the specifics of the cosmological simulation we use (\ref{method: TNG100-1}); (ii) the selection criterion for the sample of merging galaxies drawn from IllustrisTNG (\ref{method: sample selection}); (iii) the method adopted to compute the orbital parameters (\ref{method:orbital params}); (iv) the set-up for our $N$-body simulations (\ref{method: Griffin}). We present our results in Section \ref{results} and compare them with previous works in Section \ref{discussion}. We draw our conclusions in Section \ref{conclusions}. 

\section{Methods}\label{sec:methods}

\subsection{Cosmological simulation} \label{method: TNG100-1}
\label{orbits method}
IllustrisTNG is a suite of cosmological, magneto-hydrodynamic simulations with three increasingly larger physical simulation box sizes.  IllustrisTNG50, IllustrisTNG100, IllustrisTNG300 have respectively side lengths of $\sim50, 100, 300$ Mpc. Since the GWB is expected to be dominated by massive galaxies undergoing major mergers, we are interested in finding the largest possible sample of this kind of sources. This would imply choosing the biggest volume available. However, increasing the simulated volume comes at the cost of  progressively reducing the mass resolution, therefore we selected IllustrisTNG100-1  (hereafter TNG100-1),  the highest resolution simulation in the IllustrisTNG100 series. 

TNG100-1 has a volume of $110.7^3\,{\rm Mpc}^3$, with $m_{\rm{baryon}}=1.4\times 10^6 \msun$, $m_{\rm{DM}}=7.5\times 10^6 \msun$ and cosmological parameters $\Omega_{\rm{m_{tot}}}=0.3089, \Omega_{\rm{\Lambda}}=0.6911$, $\Omega_{\rm{baryon}}=0.0486$, $h=0.6774$ \citep{2016A&A...594A..13P}. The simulation uses AREPO \citep{2010MNRAS.401..791S}, a moving, unstructured-mesh hydrodynamic code, with superposed SPH (i.e. smoothed-particle hydrodynamics) gas particles \citep[e.g.][]{2010ARA&A..48..391S}, and particles that represent stars, DM (both with softening length $\varepsilon=0.74$ kpc) and SMBHs. The latter are seeded with a mass $M_{\rm{seed}}=8 \times 10^5 h^{-1}\msun$ in halos with $M_{\rm{h}}\geq5 \times 10^{10}h^{-1}\msun$. They are then allowed to accrete according to the Bondi-Hoyle accretion model \citep{1944MNRAS.104..273B} capped at the Eddington limit, and they evolve dynamically, while kept fixed at the potential minimum of their host galaxy. The simulation is initialised at $z = 127$ and evolved until $z = 0$. Data are stored in 99 snapshots that have been made publicly available online (www.tng-project.org). 

In order to easily reconstruct the merger history of the structures in the simulation, TNG provides merger trees, created using {\tt SubLink} \citep{2015MNRAS.449...49R}, an algorithm that builds merger trees at the subhalo (i.e. galactic) level. When two galaxies share the same {\tt SubLink} Descendant we consider it a merger and we define the Descendant's snapshot as the snapshot of the merger.

\subsection{Sample selection}\label{method: sample selection}
We select all galaxies in TNG100-1 with stellar mass $M_* \geq 3 \times 10^{11} \msun$ at $z=0$, thus obtaining a sample of 100 galaxies. We then follow their merger trees up to $z=2$, based on the fact that the GWB signal is dominated by sources at low redshift \citep[e.g.][]{2008MNRAS.390..192S,2022MNRAS.509.3488I}. At this stage, we retain all mergers with stellar mass ratio $q \geq 1/10$, for a total of 160 mergers. 

Data relative to the progenitor galaxies involved in the encounters are then drawn from the snapshot immediately before the merger\footnote{This is the last snapshot when the two progenitor galaxies are still distinct.}. It is worth noticing that sometimes the halo finder (i.e. the algorithm that identifies bound structures in the simulation) is not able to detect a subhalo if it is too small compared to the larger structure that it is passing through (because the density contrast is not high enough). In this eventuality, {\tt SubLink} allows some halos to skip a snapshot when finding a descendant. For this reason, there are instances in which we cannot find both progenitor galaxies in the snapshot before the merger: in this case we follow the two back in time, until both are identified in the same snapshot. If a common snapshot is not found, we remove the merger from our sample. We encountered this problem for 7 mergers, leaving us with a final sample of 153.

\subsection{Computing the orbital parameters}\label{method:orbital params}
In order to determine the orbit of the two interacting galaxies, we assume that they can be described as a Keplerian two-body system: we represent the two galaxies as two point masses of mass $M_1$ and $M_2$, positioned at their respective centres and moving with their respective bulk velocities. In order to identify the centres of the galaxies, we recursively compute the centre of mass using a shrinking sphere method on their stellar component, stopping the process when we reach a minimum of 1000 enclosed particles. The bulk velocities, on the other hand, are computed as the weighted average of the stellar velocities within the stellar half mass radius. Once these quantities are known for both galaxies, we compute the semi-major axis $a$, the orbital eccentricity $e$, the position of the pericentre $r_{\rm{peri}}$ and the position of the apocentre $r_{\rm{apo}}$ as:
\begin{equation}
    a\,=\,\left(\frac{2}{r_{\rm{rel}}}-\frac{v_{\rm{rel}}^{2}}{GM}\right)^{-1}; \textbf{ }\textbf{ } e\,=\,{\sqrt[]{1-\frac{h^{2}}{GMa}}}; 
    \label{Keplerian orbital parameters}
\end{equation}
\begin{equation}
    r_{\rm{peri}}\,=\,a\left(1-e\right); \textbf{ }\textbf{ } r_{\rm{apo}}\,=\,a\left(1+e\right)
    \label{Keplerian orbital parameters 2}
\end{equation}
where $r_{\rm{rel}}$ is the relative distance between the two centres, $v_{\rm{rel}}$ is the relative velocity, $h$ is the angular momentum per unit mass and $M=M_1+M_2$ is the total mass of the system.

\begin{figure}
    \centering
    \includegraphics[width=\columnwidth]{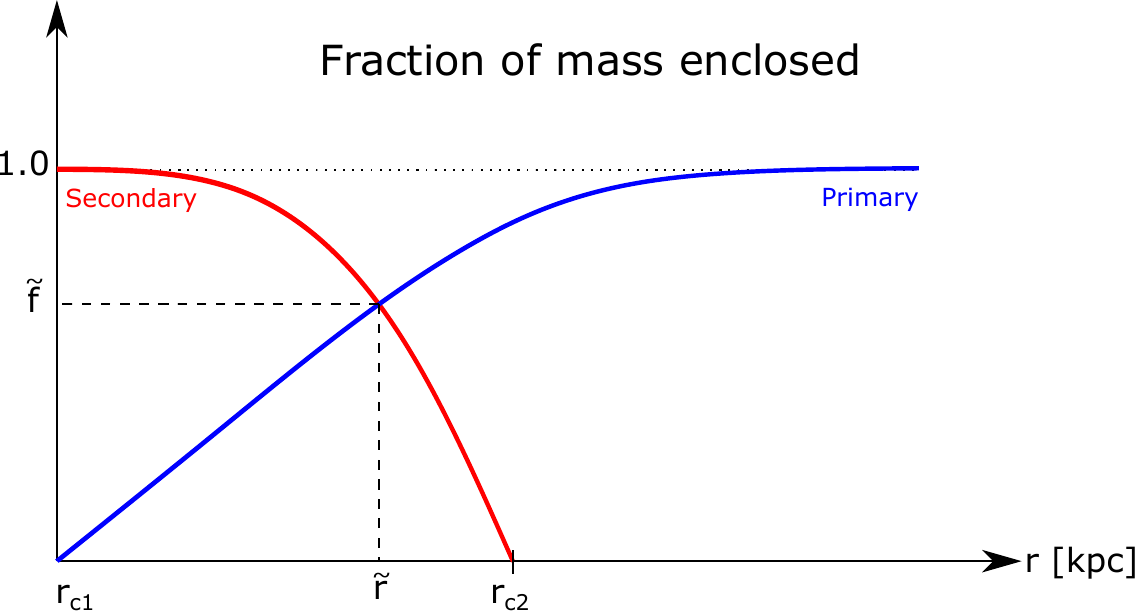}
    \caption{ Visual representation of the definition of the fraction $\tilde{f}$, where $r_{\rm{c}1}$ and $r_{\rm{c}2}$ are the centres of mass of the primary and secondary galaxy respectively. The solid line labelled "Primary" represents the enclosed fractional mass profile $f_1(\tilde{r}_1)$ of the primary galaxy, where $\tilde{r}_1=r_{\rm{c}1}+r$. For the sake of visual clarity, here the enclosed (fractional) mass profile of the secondary galaxy ($f_2(\tilde{r}_2)$, solid line labelled "Secondary") is centred in  $r_{\rm{c}2}=r_{\rm{c}1}+r_{\rm{rel}}$ (where $r_{\rm{rel}}$ is the distance between the two centres) and is mirrored with respect to the y-axis (so that the two profiles can intersect); therefore in this representation $\tilde{r}_2=r_{\rm{c}2}-r$. The point of intersection $\tilde{r}$ between the two profiles then defines $\tilde{f}$ as $\tilde{f}\equiv f_1(\tilde{r})=f_2(\tilde{r})$.}
    \label{fig:effective mass}
\end{figure}

The Keplerian approximation is justified if the two galaxies are well separated and their interaction is still negligible. In our case, however, the two DM halos are usually already overlapping at the selected time (the snapshot before the merger) and using the total mass of the system in equations (\ref{Keplerian orbital parameters}) overestimates the mass that determines the dynamics. For this reason, we replace the total mass $M$ with an \textit{effective mass} $M_{\rm{eff}}$ given by:
\begin{equation}
    M_{\rm{eff}}=\tilde{f}\;\text{(}M_{1\,}+M_{2})\,.
    \label{effective mass}
\end{equation}
To determine $\tilde{f}$ we calculate the mass enclosed within an increasing radius of both galaxies ($M_1(<\tilde{r}_1)$ and $M_2(<\tilde{r}_2)$, for the primary and secondary galaxy, respectively), starting from their centres $r_{\rm{c}_1}$ and  $r_{\rm{c}_2}$ and moving outwards. Next, we normalise the mass profiles to the total mass of each galaxy, thereby obtaining the fraction of mass enclosed at increasing radii:
\begin{equation}
 f_1(\tilde{r}_{1}) = \frac{M_1(<\tilde{r}_1)}{M_1} \text{;}\text{ }\text{ }\text{ }
 f_2(\tilde{r}_{2}) = \frac{M_2(<\tilde{r}_2)}{M_2}\,.
\end{equation}
We then partition $f_1(\tilde{r}_{1})$ and $f_2(\tilde{r}_{2})$ into discrete bins and determine at which bin the corresponding radii $\tilde{r}_1$ and $\tilde{r}_2$ satisfy the condition:   $\tilde{r}_1+\tilde{r}_2=r_{\rm{rel}}$. This allows us to find the point of intersection between the two profiles. The values of $\tilde{r}_1$ and $\tilde{r}_2$ thus obtained are then used to define $\tilde{f}$ as:
\begin{equation}
   \tilde{f}\equiv f_1(\tilde{r}_{1}) = f_2(\tilde{r}_{2})
\end{equation}
as shown in Fig.\ref{fig:effective mass}.

In the particular case of equal mass and equal size progenitors, this definition of effective mass coincides with the mass enclosed within half the separation, but it is more general and can be applied to any mass or size ratio. We tested this prescription on $N$-body simulations of equal and unequal mass progenitors and initial eccentricities of 0.9 and 0.99 and found that we can predict the distance of the first pericentre within a factor 3. 

We therefore use the effective mass to compute the orbital parameters of our merger sample. A total of 26 out of 153 orbits turn out to be unbound: this usually happens when
the most recent snapshot in which the halo-finder can find both progenitors corresponds to a significantly earlier time than the merger itself. This leaves us with a sample of 127 mergers with bound orbits. 

One could argue that, instead of introducing an effective mass, we could have drawn data from an earlier snapshot in which the progenitors are not yet overlapping. However we note that mergers (especially major ones) are extremely chaotic and minor mergers can occur between two successive snapshots.  Following the progenitors back in time is not trivial as (i) both galaxies are not guaranteed to be found in the same snapshot (as seen before), (ii) in the eventuality of minor mergers occurring in the meantime, galactic properties can change from one snapshot to the following one, (iii) if the encounter happens on a very eccentric orbit, simply moving back by one snapshot is likely to result in an unbound pair. For these reasons, we decided to adopt an effective mass in the calculation of the orbital parameters, selecting the snapshot right before the merger.

\subsection{$N$-body simulations: sub-sample selection and modelling}
\label{method: Griffin}
As our main focus is on BHBs that might significantly contribute to the GW signal in the PTA band, we  only select mergers with $z\leq1$ and $q\geq1/4$, i.e. major mergers at low redshifts.
This results in 7 mergers (hereafter Merger 1, 2, 3, 4, 5, 6, 7), whose properties are shown in Table\ref{table1 merger parameters}. We re-simulate the dynamical evolution of these mergers using  the code {\tt Griffin} \citep{2014ComAC...1....1D}, a Fast Multipole Method (FMM) $N$-body code which has been shown to obtain a distribution of force errors similar to that of a direct summation code by monitoring errors and adaptively setting expansion parameters. Encounters between stars and DM particles are modelled using the FMM technique, which (in this optimised version) scales as $\mathcal{O}\sim N^{0.87}$, while forces from interactions with SMBHs are computed via direct summation, to correctly capture their collisional nature.

\begin{figure}
    \centering
    \includegraphics[width=\columnwidth]{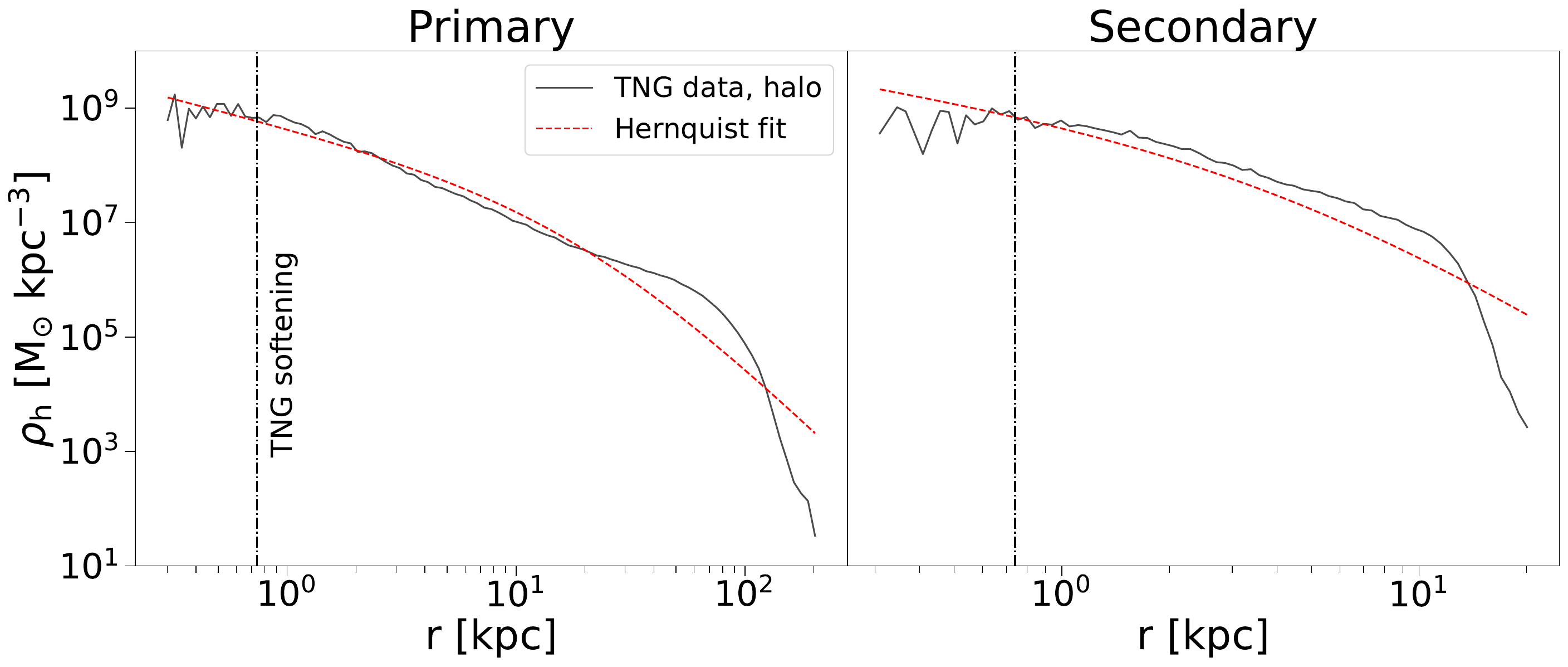}
    \includegraphics[width=1\linewidth]{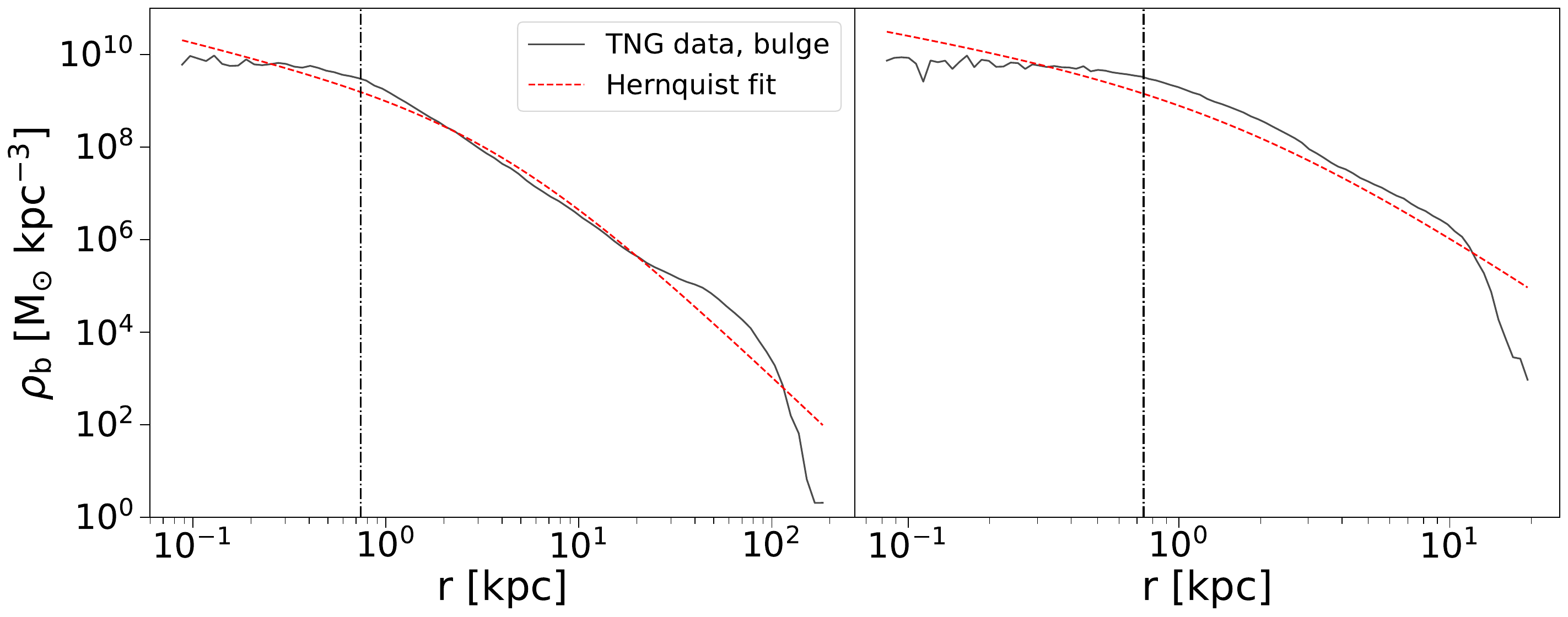}
    \caption{Density profiles of the primary and secondary galaxy in merger 7, drawn from TNG100-1 (solid line) and fitted with an Hernquist profile (dashed line). The vertical line marks the softening length in the TNG100-1 simulation. First row: DM halo density profiles. Second row: stellar bulge density profiles.}
    \label{fig:fit density profiles}
\end{figure}

\begin{table*}
\centering

\setlength{\tabcolsep}{6pt}
\begin{tabular}{|c c c c c c c c c |} 
 \hline
 ID & galaxy & $M_{\rm{h}}$[M$_{\odot}$] & $M_{\rm{b}}$[M$_{\odot}$] & $M_{\rm{BH}}$[M$_{\odot}$] &$M_{\rm{eff}}$[M$_{\odot}$] & $e$ & $r_{\rm{peri}}$[kpc] & $r_{\rm{apo}}$[kpc]    \\ [1ex] 
 \hline
 1 & primary & 1.368e+14 & 5.511e+11 & 3.713e+09 & 1.004e+12& 0.554 & 23.800 & 82.812  \\ [1ex]
 \textit{ } & secondary & 1.844e+12 & 2.269e+11 & 7.245e+08 & 1.587e+09& \textit{ } & \textit{ } & \textit{ }   \\ [1ex]
 \hline 
 2 & primary & 1.091e+13 & 1.693e+11 & 3.205e+08 & 4.112e+11& 0.994 & 0.304 & 94.713   \\[1ex]
 \textit{ } & secondary & 4.226e+10 & 4.461e+10 & 3.716e+08 & 3.010e+09& \textit{ } & \textit{ } & \textit{ }   \\[1ex]
 \hline 
 3 & primary & 1.371e+13 & 2.235e+11 & 9.042e+08 & 1.157e+12& 0.995 & 0.163 & 68.805   \\[1ex]
 \textit{ } & secondary & 8.943e+10 & 6.086e+10 & 2.145e+08 & 1.119e+10& \textit{ } & \textit{ } & \textit{ }   \\[1ex]
 \hline 
 4 & primary & 1.272e+13 & 9.121e+10 & 5.704e+08 & 6.431e+11& 0.995 & 0.123 & 46.326   \\[1ex]
 \textit{ } & secondary & 5.895e+10 & 7.131e+10 & 8.220e+07 & 6.009e+09& \textit{ } & \textit{ } & \textit{ }   \\[1ex]
 \hline 
 5 & primary & 1.416e+13 & 4.031e+11 & 1.086e+09 & 2.106e+12& 0.976 & 0.932 & 78.275   \\ [1ex]
 \textit{ } & secondary & 2.936e+11 & 1.257e+11 & 4.573e+08 & 5.848e+10& \textit{ } & \textit{ } & \textit{ }  \\ [1ex]
 \hline 
 6 & primary & 1.290e+13 & 3.162e+11 & 1.748e+09 & 1.212e+12& 0.987 & 0.299 & 46.363   \\ [1ex]
 \textit{ } & secondary & 6.928e+10 & 1.056e+11 & 1.347e+09 & 1.451e+10& \textit{ } & \textit{ } & \textit{ }    \\ [1ex]
 \hline 
 7 & primary & 2.222e+12 & 2.342e+11 & 7.247e+08 & 6.468e+11& 0.886 & 3.640 & 60.122   \\ [1ex]
 \textit{ } & secondary & 1.240e+11 & 8.942e+10 & 7.305e+08 & 5.499e+10& \textit{ } & \textit{ } & \textit{ }   \\ [1ex]
 
 \hline
\end{tabular}
\caption{Properties of the 7 selected mergers: merger identifier, primary/secondary galaxy, halo mass $M_{\rm{h}}$, bulge mass $M_{\rm{b}}$ and SMBH mass $M_{\rm{BH}}$, taken directly from TNG100-1, effective mass $M_{\rm{eff}}$ used to compute the initial galaxy orbits (see eq. \ref{effective mass}), and resulting orbital eccentricity $e$, position of the pericentre $r_{\rm{peri}}$ and position of the apocentre $r_{\rm{apo}}$ computed as per eq. \ref{Keplerian orbital parameters}, \ref{Keplerian orbital parameters 2} using $M_{\rm{eff}}$ and TNG100-1 data.}
\label{table1 merger parameters}
\end{table*}
To set the initial conditions for our {\tt Griffin} simulations, we re-model the progenitor galaxies based on data drawn from TNG100-1 using AGAMA \citep{2019MNRAS.482.1525V}, an action-based galaxy modelling software that can generate a potential for each galactic component (i.e. stellar bulge, DM halo and SMBH). The stellar bulge and the DM halo follow an Hernquist profile \citep{1990ApJ...356..359H}:

\begin{equation}
    \rho(r)=\frac{M}{2\pi a^3}\frac{a}{r}\frac{1}{(1+r/a)^3}
    \label{hernquist profile}
\end{equation}
where $M$ is the total mass and $a$ is the scale radius. This profile represents a good fit to the galaxies in TNG100-1, as shown in Fig.\ref{fig:fit density profiles}, for both bulge and halo components. The fits also provide best values for the scale radius.

We note that the density profiles from TNG100-1 are not reliable below $\sim1\kpc$, corresponding to the simulation's resolution limit, implying that the central slope ($\gamma$) of the profile is not constrained by the data. The Hernquist profile belongs to the one-parameter family of Dehnen profiles \citep{1993MNRAS.265..250D} which differ only for the value of $\gamma$:
\begin{equation}
    \rho(r)=\frac{(3-\gamma)M}{4\pi a^3}\frac{a^\gamma}{r^\gamma}\frac{1}{(1+r/a)^{4-\gamma}}
    \label{dehnen profile}
\end{equation}
with $0\leq\gamma <3$ and $\gamma=1$ corresponding to the Hernquist model. While any value of the $\gamma$ slope could in principle fit the data and be adopted, we have selected the Hernquist profile as it is unlikely that the progenitor galaxies will have already undergone one or more major mergers, producing an extremely flat core, and a shallow cusp appears a likely common outcome.

We model the primary galaxy using $N=10^6$ particles, half of which are assigned to the stellar bulge and half to the DM halo. We increase the resolution of the central part of the galaxy with the mass refinement scheme of \cite{2024MNRAS.529.2150A}. The scheme divides particles into several radial shells and over-samples particles in the central zone at the expense of those in the outermost zones. This is achieved by increasing the total number of particles up to a factor $10$, with the added particles being retained in the central shell. The scheme then progressively removes particles moving outwards and proportionally increases the mass of the remaining ones. In this way, the total mass and density profile are preserved.

The number of particles in the secondary galaxy, on the other hand, is set according to the following requirements: (i) same particle masses, for both stars and DM particles, in the primary and secondary galaxy; (ii) all mass ratios to be preserved, namely the one between the bulge and the halo and the one between the primary and the secondary. We then apply the mass refinement scheme to the secondary as well. Particle numbers for all components of each simulation are listed in Table \ref{table:N particles}.

\begin{table}
\centering

\setlength{\tabcolsep}{4pt}
\begin{tabular}{|c c c c c|} 
 \hline 
 ID & galaxy & $N_{\rm{tot}}$ & $N_{\rm{b}}$ & $N_{\rm{h}}$ \\ [1ex] 
 \hline
 1 & Primary & 1054436 & 527218 & 527218 \\ [1ex]
 & Secondary & 224175 & 217068 & 7107 \\ [1ex]
 \hline 
 2 & Primary & 1054436 & 527218 & 527218 \\[1ex]
 & Secondary & 140965 & 138921 & 2044 \\[1ex]
 \hline 
 3 & Primary & 1054436 & 527218 & 527218 \\[1ex]
 & Secondary & 147002 & 143563 & 3439 \\[1ex]
 \hline 
 4 & Primary & 1054436 & 527218 & 527218 \\[1ex]
 & Secondary & 414637 & 412192 & 2445 \\[1ex]
 \hline 
 5 & Primary & 1054436 & 527218 & 527218 \\ [1ex]
 & Secondary & 175337 & 164405 & 10932 \\ [1ex]
 \hline 
 6 & Primary & 1054436 & 527218 & 527218 \\ [1ex]
 & Secondary & 178905 & 176073 & 2832 \\ [1ex]
 \hline 
 7 & Primary & 1054436 & 527218 & 527218 \\ [1ex]
 & Secondary & 230721 & 201299 & 29422 \\ [1ex]
 
 \hline
\end{tabular}
\caption{For each merger we report the total number of particles used ($N_{\rm{tot}}$), the number of particles assigned to the stellar bulge ($N_{\rm{b}}$) and to the DM halo ($N_{\rm{h}}$), both for the primary and secondary galaxy, after mass refinement. 
}
\label{table:N particles}
\end{table}

We caution that as a result of the application of a mass refinement scheme to increase central resolution, the mass ratio between DM and stellar particles will increase. This can in principle lead to mass segregation of halo particles into the bulge on timescales that are comparable with that of the galactic merger. In order to mitigate this effect, we increase the softening length of massive particles, exploiting Griffin's option to define an individual softening for each particle in the simulation. The softening values for standard particles (stars and DM) are chosen as follows:
\begin{equation}
    \varepsilon_{\rm{std}} = \alpha m_{\rm{std}}^{1/3}
\end{equation}
where $m_{\rm std}$ is the mass of the particle and $\alpha$ is the proportionality constant given by
\begin{equation}
    \alpha = \frac{\varepsilon_{0,\rm{std}}}{m_{\rm{h,sh1}}^{1/3}}
\end{equation}
where $\varepsilon_{0,\rm{std}}= 30$ pc and $m_{\rm{h,sh1}}$ is the mass of halo particles in the innermost shell in code units. Similarly, the softening of the SMBHs is given by: 
\begin{equation}
    \varepsilon_{\rm{BH}}=\varepsilon_{0,\rm{BH}}m_{\rm{BH}}^{1/3}
    \label{soft_bh}
\end{equation}
where  $\varepsilon_{0,\rm{BH}}=3$ pc \footnote{This value is assigned just before the hardening phase of the BHB begins. During the dynamical friction phase we usually set  $\varepsilon_{0,\rm{BH}}$ to a higher value ($6$ or $10\pc$) in order to reduce the computational time.} and $m_{\rm{BH}}$ is the SMBH mass.  

\section{Results}\label{results}
\subsection{Statistics on TNG100-1 mergers}\label{results: statistics}
\begin{figure}
    \centering
    \includegraphics[width=\columnwidth]{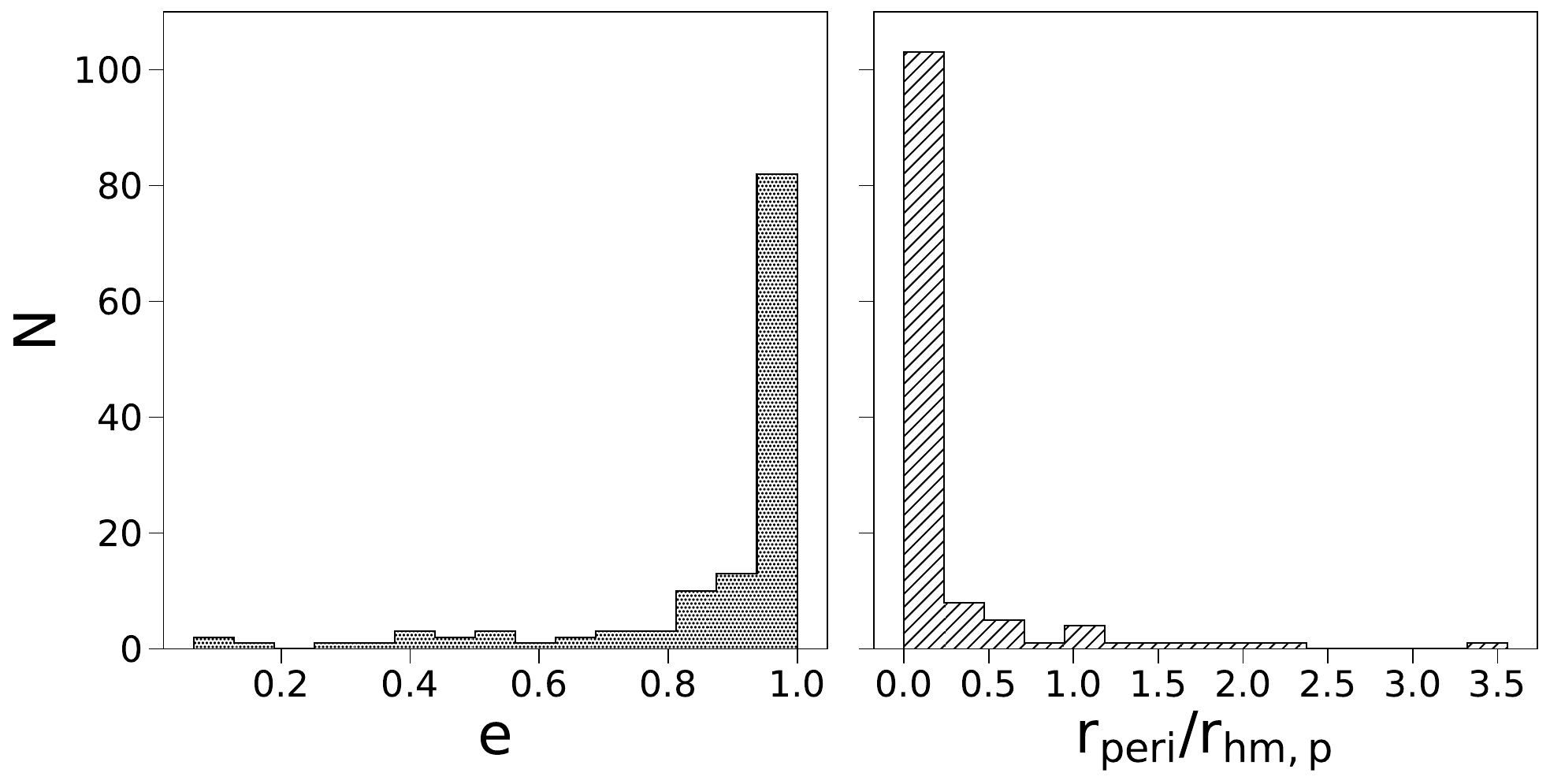}
    \caption{Left panel: distribution of the orbital eccentricity of the 127 galactic mergers in our sample. Right panel: distribution of the position of the pericentre, normalised to the half mass radius of the primary galaxy (as reported in TNG100-1). The primary galaxy is defined as the more massive of the two progenitors.}
    \label{fig:histos}
\end{figure}

\begin{figure}
    \centering
    \includegraphics[width=\columnwidth]{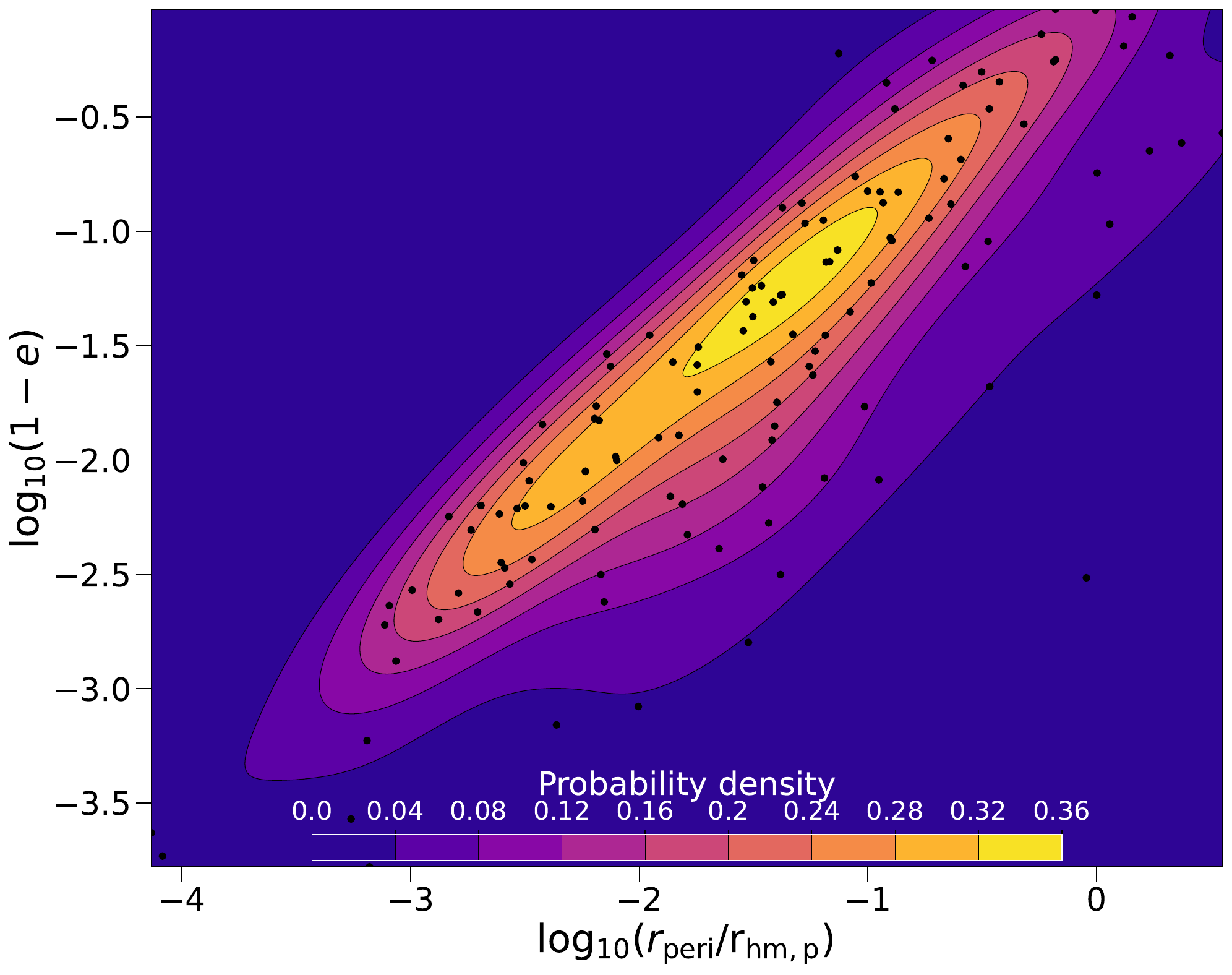}
    \caption{Eccentricity-pericentre parameter space: each circle represents one of the 127 mergers in our sample. The colour map highlights regions of increasing probability density. The highest density region corresponds to the peak of the distribution, with $e\sim0.97$ and $r_{\rm{peri}}/r_{\rm{hm,p}}\sim0.04$.}
    \label{fig:contour-plot}
\end{figure}

We compute the orbital parameters of the BHBs that form in all the 127 mergers belonging to our sample, as described in Sec. \ref{method:orbital params}. Our main interest is to assess what the typical eccentricities of these orbits are and how close the pericentric passages are, since (i) we expect a correlation between the initial eccentricity of the merger and the eccentricity of the binary at formation \citep{2022MNRAS.511.4753G} and (ii) if the secondary galaxy penetrates well within the primary, we expect dynamical friction to be more efficient and the merger to proceed faster. The distributions of eccentricity and pericentre distance (normalised to the half-mass radius of the primary galaxy) are shown in Fig.\ref{fig:histos}. The eccentricity distribution clearly peaks at high values, with a mean of $\overline{e}=0.88$ and a median of $e_{\rm{m}}=0.97$\footnote{The median value is more relevant in this case, given the asymmetry of the distribution}. 
Furthermore, the majority of the encounters have a pericentric passage well within the half-mass radius of the primary galaxy (r$_{\rm{hm,p}}$), with a median value of 
$\sim 0.04 r_{\rm hm,p} $. In Fig.\ref{fig:contour-plot} we show these results in the eccentricity-pericentre parameter space: each circle in the plot represents one merger and the colour map defines regions of increasing probability density, computed using a kernel density estimation (KDE) function. 

\subsection{Evolution through binary formation and hardening}\label{results: Griffin}
The 7 mergers belonging to our sub-sample (as defined in Sec. \ref{method: Griffin}) are followed in their evolution using the {\tt Griffin} code from the onset of the galactic merger to BHB formation and hardening. The progenitor galaxies are placed at the apocentre of their orbit (computed as described in Sec. \ref{method:orbital params}) and evolved through the dynamical friction, binary formation and hardening phases.  

\begin{figure}
    \centering
    \includegraphics[width=\columnwidth]{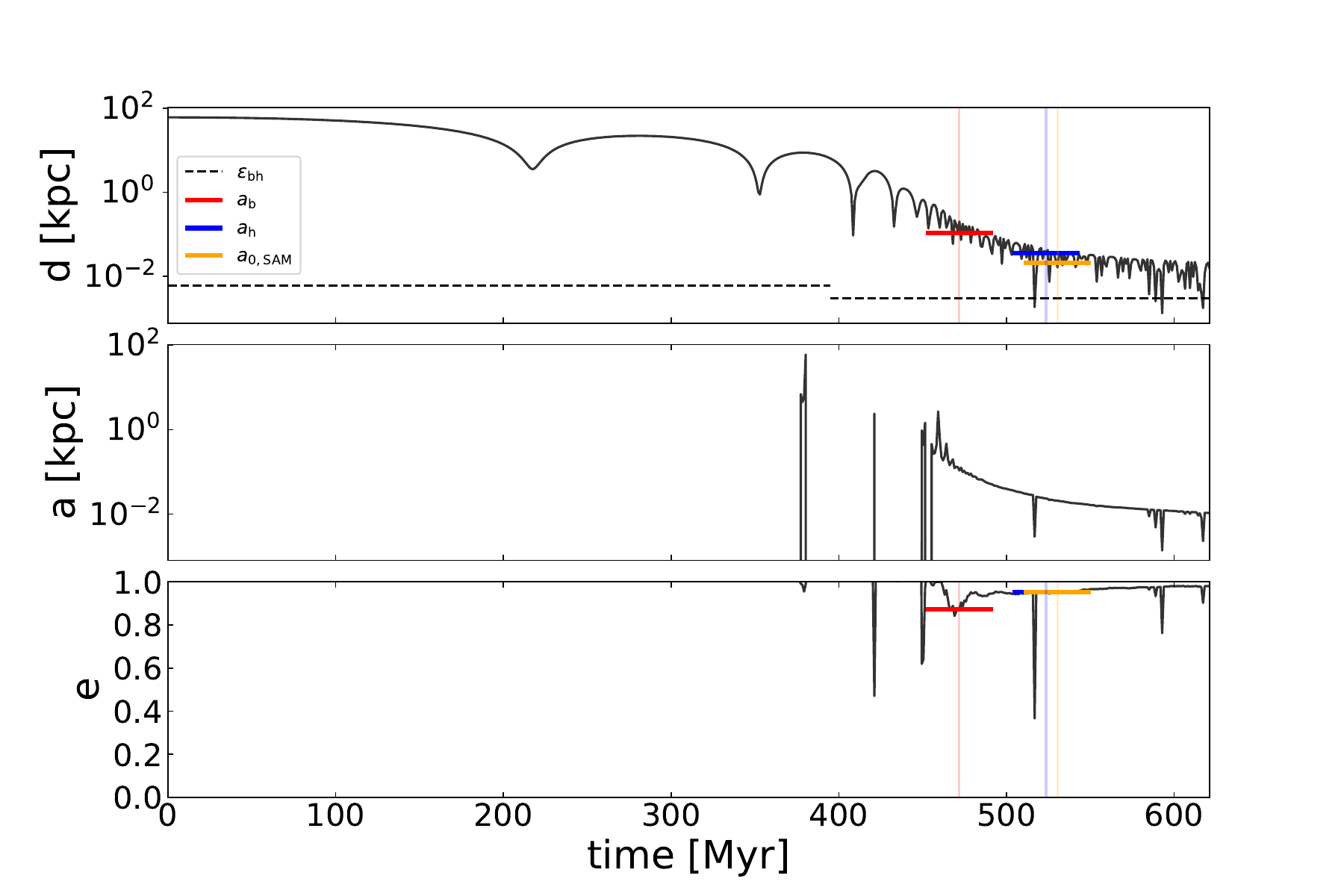}
    \caption{Evolution of the distance between the black holes (top panel), Keplerian  semi-major axis (middle panel) and eccentricity (bottom panel) in merger 7.
    The horizontal dashed lines represent the softening parameter $\varepsilon_{0,\rm{BH}}$ (see eq.\ref{soft_bh}) used for the black holes.  The dotted vertical lines represent, from left to right, the time $t_{\rm{b}}$ of binary formation, the time $t_{\rm{h}}$ when the hard-binary separation is reached, and the time $t_{\rm{0,SAM}}$ when the semi-analytical modelling is started. The binary forms with very large eccentricity.}
    \label{fig:soft7_params}
\end{figure}

\begin{figure}
    \centering
    \includegraphics[width=\columnwidth]{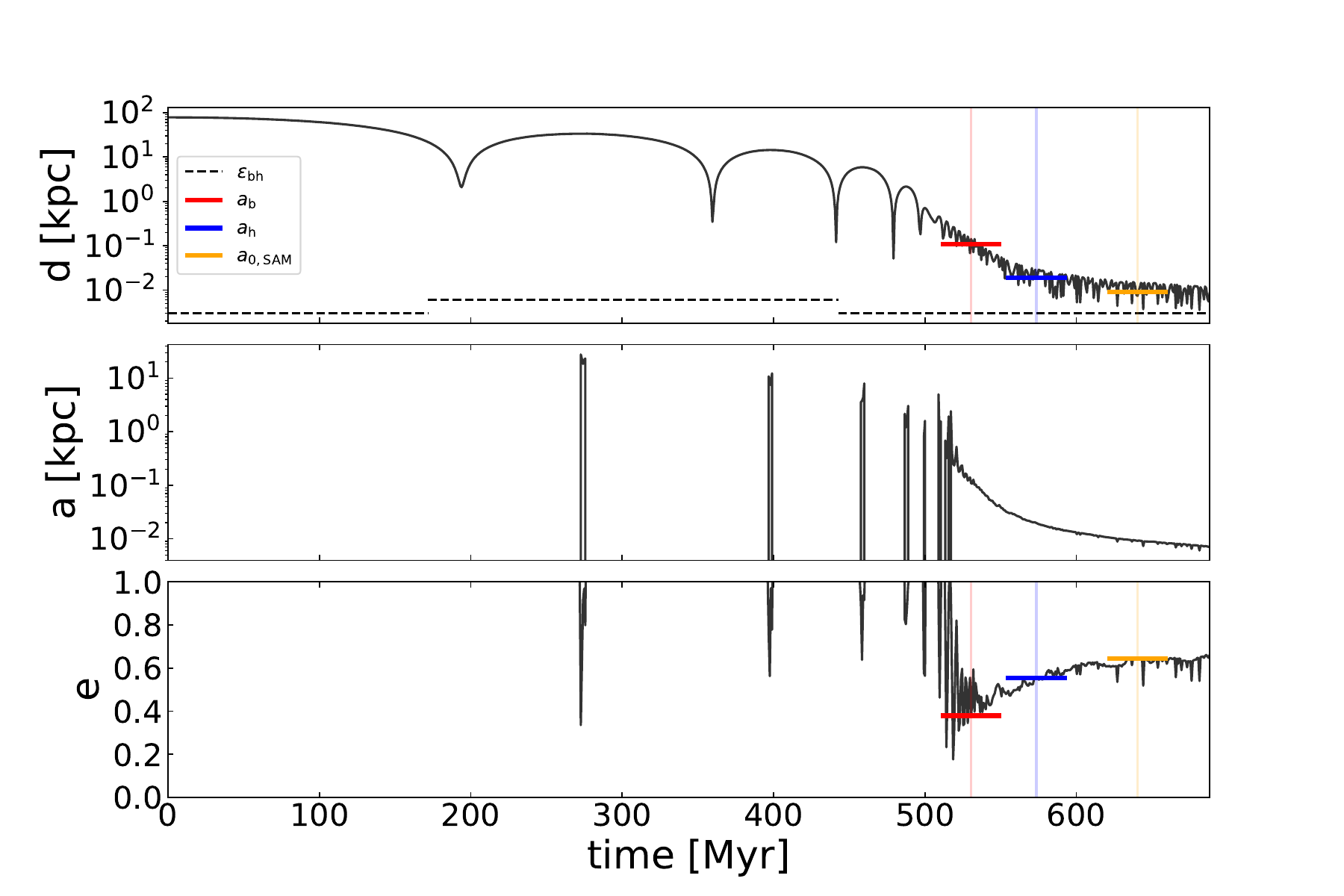}
    \caption{Same as Fig.\ref{fig:soft7_params} but for merger 5; in this merger the eccentricity after binary formation is remarkably lower.}
    \label{fig:soft5_params}
\end{figure}

Fig. \ref{fig:soft7_params} and \ref{fig:soft5_params} show the evolution of two of the simulated mergers\footnote{These mergers are chosen as representative of two different behaviours found in our complete sub-sample
%. Figures relative to the remaining mergers are shown in Appendix \ref{appendix A}.
}.  
We recognise a first, slower phase in the evolution corresponding to the galactic merger and dynamical friction phase, when the two SMBHs are still unbound and inspiralling towards the centre of the merger remnant.  
The black holes then enter a binding phase, in which they oscillate between a bound and an unbound state, clearly visible in the plots of the Keplerian semi-major axis and eccentricity. 
The binary eventually settles on a bound orbit and enters the gravitational slingshot phase. Here the separation between the SMBHs shrinks very quickly, as the binary hardens due to the ejection of stars following a close three-body encounter. 
Once all the stars initially populating the binary's loss cone have been ejected, the evolution slows down again and relies on interactions with stars refilling the loss cone due to angular momentum diffusion. As expected, the eccentricity tends to increase after binary formation, due to the three-body interactions with stars \citep[e.g.][]{2006ApJ...651..392S}.

Critical times and their respective values of distance and eccentricity are marked in different colours in Fig.\ref{fig:soft7_params} and \ref{fig:soft5_params} from left to right: (i) the binding time ($t_{\rm{b}}$), defined empirically as the time at which the polynomial fit of the eccentricity evolution reaches the minimum. 
This identifies, in an arbitrary way, the point in the binary evolution when the Keplerian orbital parameters become well defined, and excludes the initial chaotic phase; (ii) the hard-binary time ($t_{\rm{h}}$), defined as the time when the semi-major axis of the binary reaches the hard-binary separation $a_h = \frac{q}{(1+q)^2}\frac{r_{\rm{m}}}{4}$ \citep{2006RPPh...69.2513M}, where $q$ is the mass ratio and $r_{\rm{m}}$ is the radius containing a mass in stars equal to twice the mass of the primary; (iii) the time $t_{0,\rm{SAM}}$ where we start the semi-analytical modelling of the evolution (see Sec. \ref{results:sam}).

\begin{figure}
    \centering
    \includegraphics[width=1\linewidth]{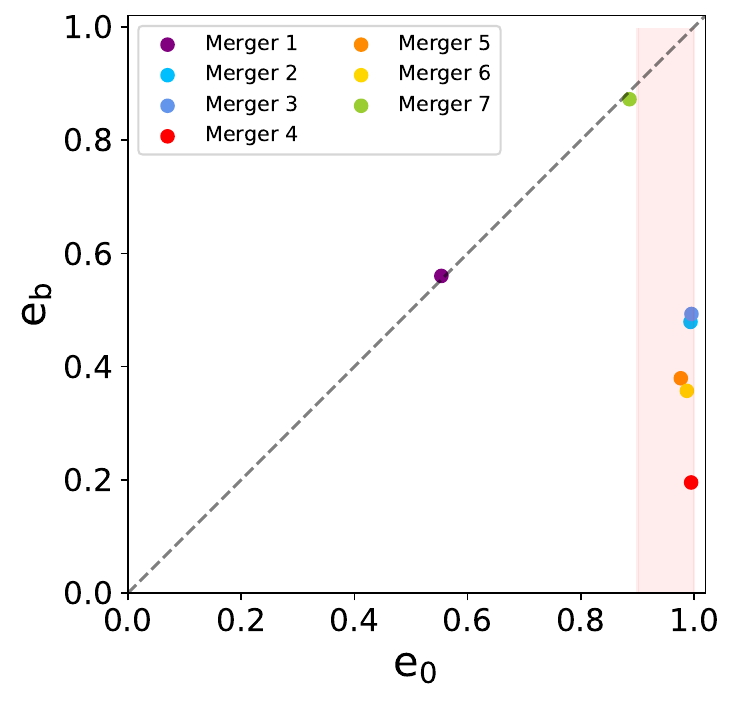}
    \caption{Relation between the initial eccentricity of the galactic merger $e_0$ and the eccentricity of the binary at binding $e_{\rm{b}}$. We highlight the region with $e_0>0.9$ with a shaded area: above this threshold more circular binaries tend to form, with a significant spread in eccentricity, breaking the expected correlation between $e_0$ and $e_{\rm{b}}$}.
    \label{fig:correlation}
\end{figure}

Interestingly, we find that the eccentricity with which BHBs form depends on the initial eccentricity $e_0$ of the galactic merger, but only up to a threshold value. Mergers with high initial eccentricities tend to form highly eccentric binaries \citep[in agreement with the expected correlation observed in][] {2022MNRAS.511.4753G}, as long as $e_0\lesssim0.9$; mergers with initial eccentricities $e_0\gtrsim0.9$ tend to form more circular binaries, breaking the aforementioned correlation (see Fig. \ref{fig:correlation} ). We caution that the sample size used in our study is small and therefore the results are subject to low number statistics. However, we would like to emphasise that (i) we can reproduce the correlation observed in \cite{2022MNRAS.511.4753G} for comparable eccentricity values (i.e. for the two mergers with e$_0\leq 0.9$); (ii) none of the five mergers with e$_0>0.9$ follows the expected correlation. This suggests that further investigation is required to gain a better understanding of these highly eccentric systems.

%\section{Completing the evolution with a Semi-Analytical Model}\label{sam}
\subsection{Evolution through Gravitational Wave emission and coalescence}\label{results:sam}
As previously mentioned, {\tt Griffin} simulates three-body interactions between the binary and individual stars via direct summation; this inevitably implies that the computational cost of modelling the BHB through the whole hardening phase till GW emission commences
is unsustainable.  We therefore model the late evolution of the binaries to coalescence by means of a semi-analytical model (SAM). The SAM combines the effects of both stellar interactions (referred to by the subscript $\star$) and GW emission (referred to by the subscript GW), with the latter becoming dominant at later times. The rate of change of the semi-major axis and the eccentricity of the binary can be modelled as:
\begin{align}
    \frac{da}{dt} &= \left. \frac{da}{dt} \right|_\star + \left.\frac{da}{dt}\right|_{\rm GW}\label{dadt}\\
    \frac{de}{dt} &= \left.\frac{de}{dt}\right|_\star + \left.\frac{de}{dt}\right|_{\rm GW}\label{dedt}\,.
\end{align}
The evolution due to stellar interactions can be described as \citep{1996NewA....1...35Q}:
\begin{align}
    \left. \frac{da}{dt} \right|_\star  = -a^2\frac{HG\rho}{\sigma}\label{da/dt_star}\\ \left. \frac{de}{dt} \right|_\star = a \frac{HKG\rho}{\sigma}\label{de/dt_star}
\end{align}
where $\rho$ and $\sigma$ are respectively the stellar density and velocity dispersion within the radius of influence $r_{\rm{inf}}$ of the binary\footnote{$r_{\rm{inf}}$ is usually defined as the radius at which the stellar mass enclosed within the orbit of the binary is twice the mass of the BHB.}, while $H$ and $K$ represent the dimensionless hardening rate and eccentricity growth rate, respectively. They depend on the binary's mass ratio, eccentricity and separation and can be derived through three-body scattering experiments of the ejection of background stars by the BHB.
We adopt the tabulated parameters given in \citet{2006ApJ...651..392S}, interpolating as required for our merger configurations.

The evolution due to GW emission can be modelled using Peters' equations \citep{1964PhRv..136.1224P}:
\begin{align}
    \left. \frac{da}{dt} \right|_{\rm GW}  &= -\frac{64 G^3}{5c^5}\frac{M_1M_2M}{a^3(1-e^2)^{7/2}}\left(1+\frac{73}{24}e^2+\frac{37}{96}e^4\right)\label{dadt_gw} \\
    \left. \frac{de}{dt} \right|_{\rm GW} &=  -\frac{304 G^3}{15c^5}\frac{M_1M_2M}{a^4(1-e^2)^{5/2}}\left(e+\frac{121}{304}e^3\right)\label{dedt_gw}
\end{align}
where $G$ is the gravitational constant, $c$ is the speed of light in vacuum, $M_1$ and $M_2$ are the masses of the primary and secondary SMBH, respectively, and $M = (M_1 + M_2)$ is the total binary mass.  Finally, eq. \ref{dadt} and \ref{dedt} are integrated using the Euler method until one of the following requirements is no longer fulfilled: (i) $a>0$, (ii) $0\leq e<1$, (iii) $t<10^{10}\gyr$.

For each model, the SAM starts at a time $t_{0,\rm{SAM}}$ when the binary is already hard: $t_{0,\rm{SAM}}>t_{\rm{h}}$. We select the snapshot in the {\tt Griffin} simulation closest to this time and extract values for the semi-major axis and the eccentricity of the BHB ($a_{0,\rm{SAM}}$ and $e_{0,\rm{SAM}}$ respectively, listed in Table \ref{table:sam}), which become the initial conditions for the model. We compute $\rho$ and $\sigma$ at the radius of influence of the binary at the same time $t_{0,\rm{SAM}}$.

\begin{figure}
    \centering
    \includegraphics[width=\columnwidth]{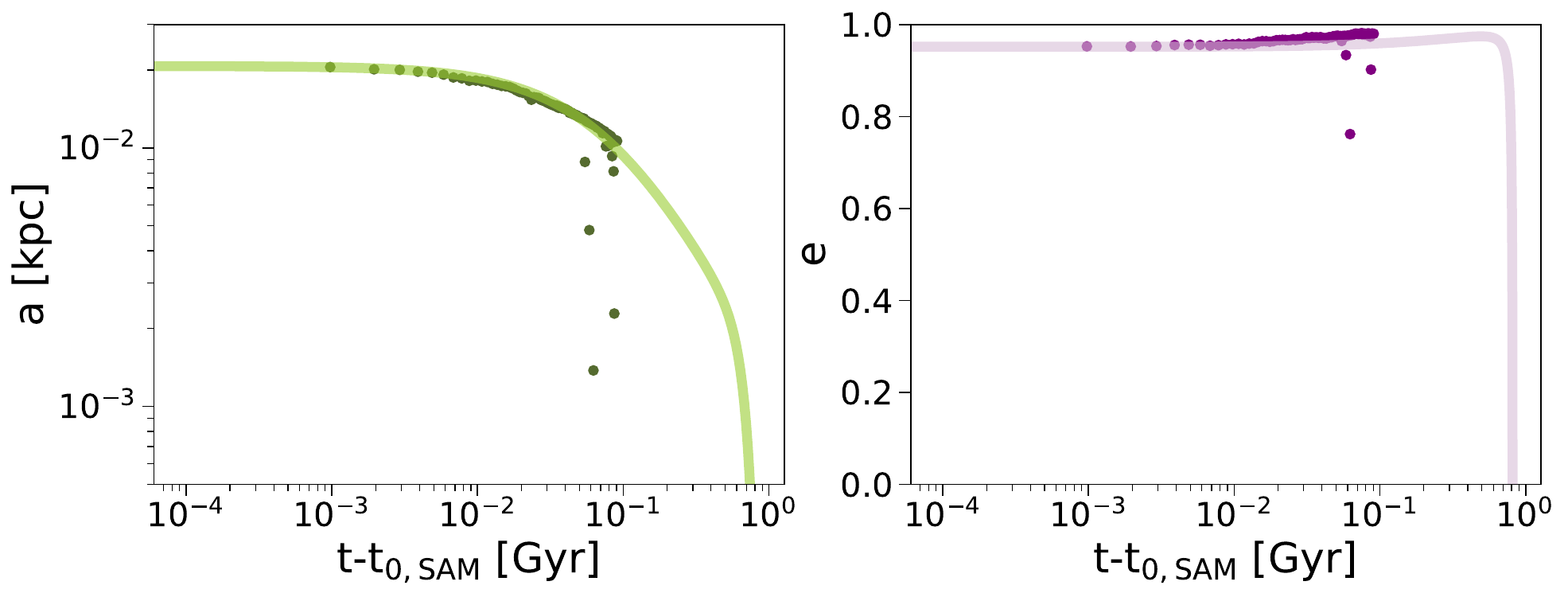}
    \caption{Evolution of the semi-major axis (left) and eccentricity of the BHB (right) in Merger 7 computed with the SAM through the hardening and GW emission phase to coalescence (solid lines). The circles represent the orbital elements calculated from the last few snapshots of the $N$-body simulations. The time on the $x$-axis is measured from the beginning of the SAM at $t_{0,\rm{SAM}}=0.53\gyr$.}
        \label{fig:soft7_SAM}
\end{figure}

\begin{figure}
    \centering
    \includegraphics[width=\columnwidth]{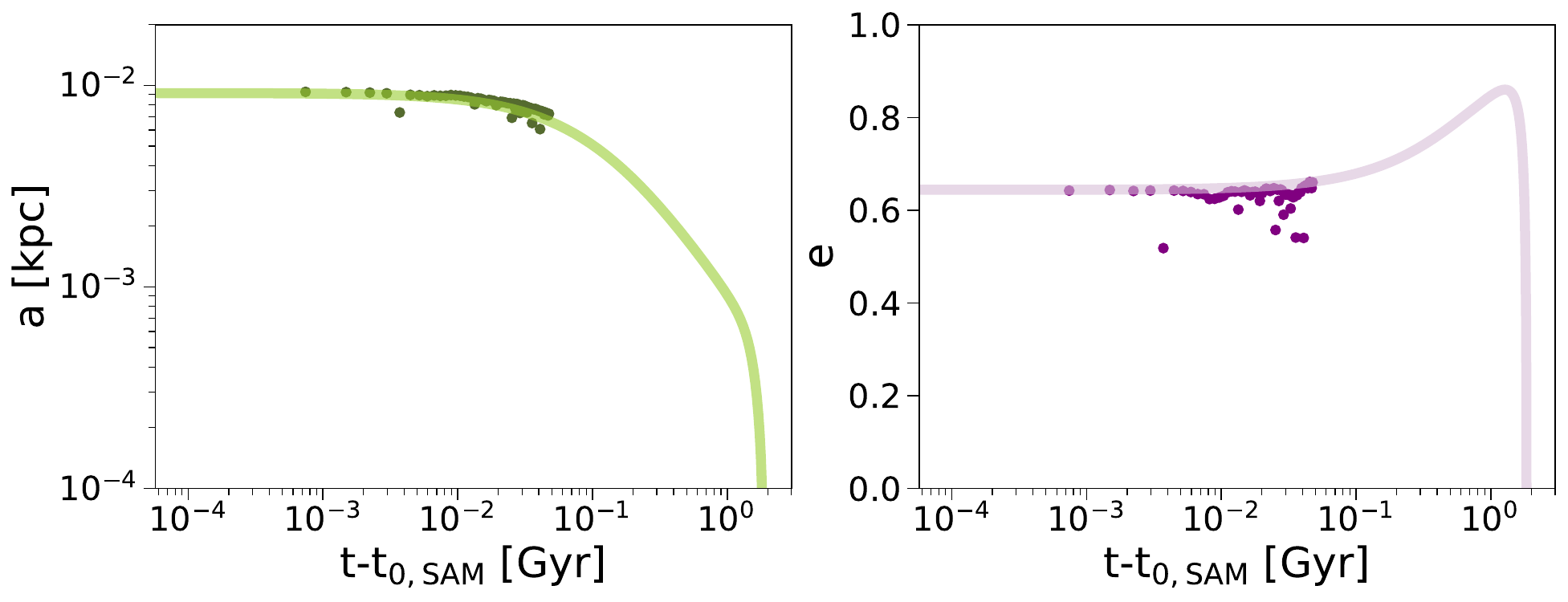}
    \caption{Same as Fig.\ref{fig:soft7_SAM} but for Merger 5. For this merger  $t_{0,\rm{SAM}}=0.64\gyr$.}
    \label{fig:soft5_SAM}
\end{figure}

 We ensure that the SAM reproduces the $N$-body evolution by continuing the {\tt Griffin} simulations for a few additional snapshots and comparing with the predictions of the SAM. We find that, in order to reproduce the evolution of the semi-major axis, we need to lower the tabulated values of $H$ reported in \citep{2006ApJ...651..392S} by $\sim 20-40\%$\footnote{$H$ is lowered by $20\%$ in Merger 1, 2, 3 and by $40\%$ in Merger 4, 5, 6, 7.}. One possible explanation for the high values of $H$ obtained in scattering experiments is that they assume that the system is always in the full loss-cone regime, which is not necessarily true in $N$-body simulations. Furthermore, we increase the values of $K$ by a factor $1.5$, based on a follow-up study on scattering experiments \citep{2019ApJ...878...17R}, that points out an error in the original calculation of $K$ in \cite{2006ApJ...651..392S}. Fig. \ref{fig:soft7_SAM} and \ref{fig:soft5_SAM} show the evolution of the orbital parameters for both Merger 7 and Merger 5 obtained with the SAM, down to coalescence, compared with the data taken from the last snapshots in the $N$-body simulation. Agreement with the {\tt Griffin} evolution is very good, despite the noise characteristic of the $N$-body data, implying that the extrapolation to late times and the estimate of the coalescence time provided by the SAM are reliable.

\begin{table}
\centering

\setlength{\tabcolsep}{0.01\linewidth}
\begin{tabular}{|c |c c c|} 
 \hline
 ID & $t_{0,\rm{SAM}}$ [Gyr] & $a_{0,\rm{SAM}}$ [kpc] & $e_{0,\rm{SAM}}$\\ [1ex]
 \hline 
 1 & 0.73 & 0.04 & 0.82 \\[1ex]

 2 & 2.3 & 0.02 & 0.73 \\[1ex]

 3 & 1.28 & 0.01 & 0.57  \\[1ex]

 4 & 0.57 & 0.01 & 0.66 \\[1ex]

 5 & 0.64 & 0.01 & 0.64 \\ [1ex]
 
 6 & 0.44 & 0.01 & 0.44 \\ [1ex]

 7 & 0.53 & 0.02 & 0.95 \\ [1ex]

 \hline
\end{tabular}
\caption{Initial conditions used for the semi-analytical models.}
\label{table:sam}
\end{table}

Table \ref{table:sum new complete} lists merger parameters at critical times in the evolution: (i) at the beginning of the $N$-body simulation (referred to by the subscript "0"); (ii) at binding time (referred to by the subscript "b"); (iii) at the time corresponding to the hard-binary separation (referred to by the subscript "h"); (iv) at coalescence. The coalescence time $t_{\rm{coal}}$ is the total time elapsed between the start of the $N$-body simulation and the end of the SAM evolution.

We note that Merger 2 and 3 present a steep increase of orbital eccentricity in the {\tt Griffin} simulations after binary formation that cannot be reproduced via the SAM, unless the $K$ parameter is further increased with respect to the value tabulated in \cite{2006ApJ...651..392S}. For these two mergers, we therefore report both the coalescence time obtained increasing $K$ by a factor 1.5 (as for the other mergers) and the coalescence time predicted increasing $K$ to match the $N$-body data (a factor 10 for Merger 2 and a factor 6 for Merger 3).

\begin{figure}
   \centering
   \includegraphics[width=\columnwidth]{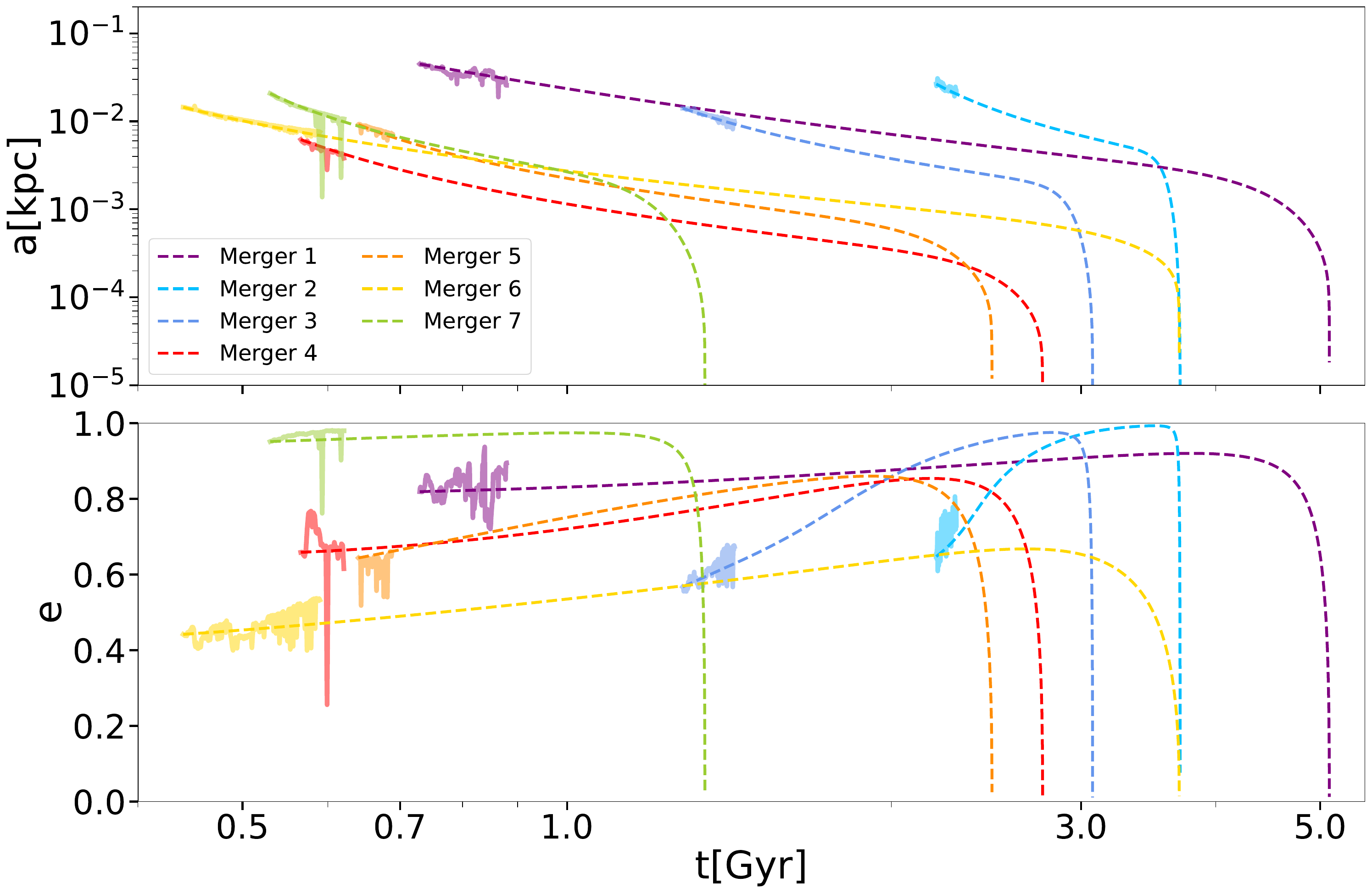}
    \caption{Evolution of the semi-major axis (top panel) and eccentricity (bottom panel) of the BHBs formed in all of the selected mergers as determined by the SAM (dashed lines), overlapped with the $N$-body evolution from time $t_{0,\rm{SAM}}$ (solid lines). For Mergers 2 and 3 we plot the evolution obtained with increased values of $K$, to match the steep increase in eccentricity observed in the $N$-body data.}
    \label{fig: sum new}
   \end{figure}

We converted the TNG100-1 redshift of all mergers to the respective look-back time T$_{\rm{lb}}$ and compared it with the coalescence time predicted by the SAM: with the exception of Merger 1, all mergers satisfy the condition T$_{\rm{lb}}>t_{\rm{c}}$. This is true also for Merger 2 and 3, if we adopt the SAM tuned on the $N$-body data described above, meaning that the BHBs are observable GW sources.
Fig. \ref{fig: sum new} shows the complete evolution of the orbital elements of all the mergers as determined by the SAM, overlapped with the $N$-body evolution from time $t_{0,\rm{SAM}}$. The strong dependence of the time spent in the GW phase on the hardening and eccentricity growth rate is evident, with a difference of several Gyr in the coalescence time of different models.

\begin{figure}
    \centering
    \includegraphics[width=\columnwidth]{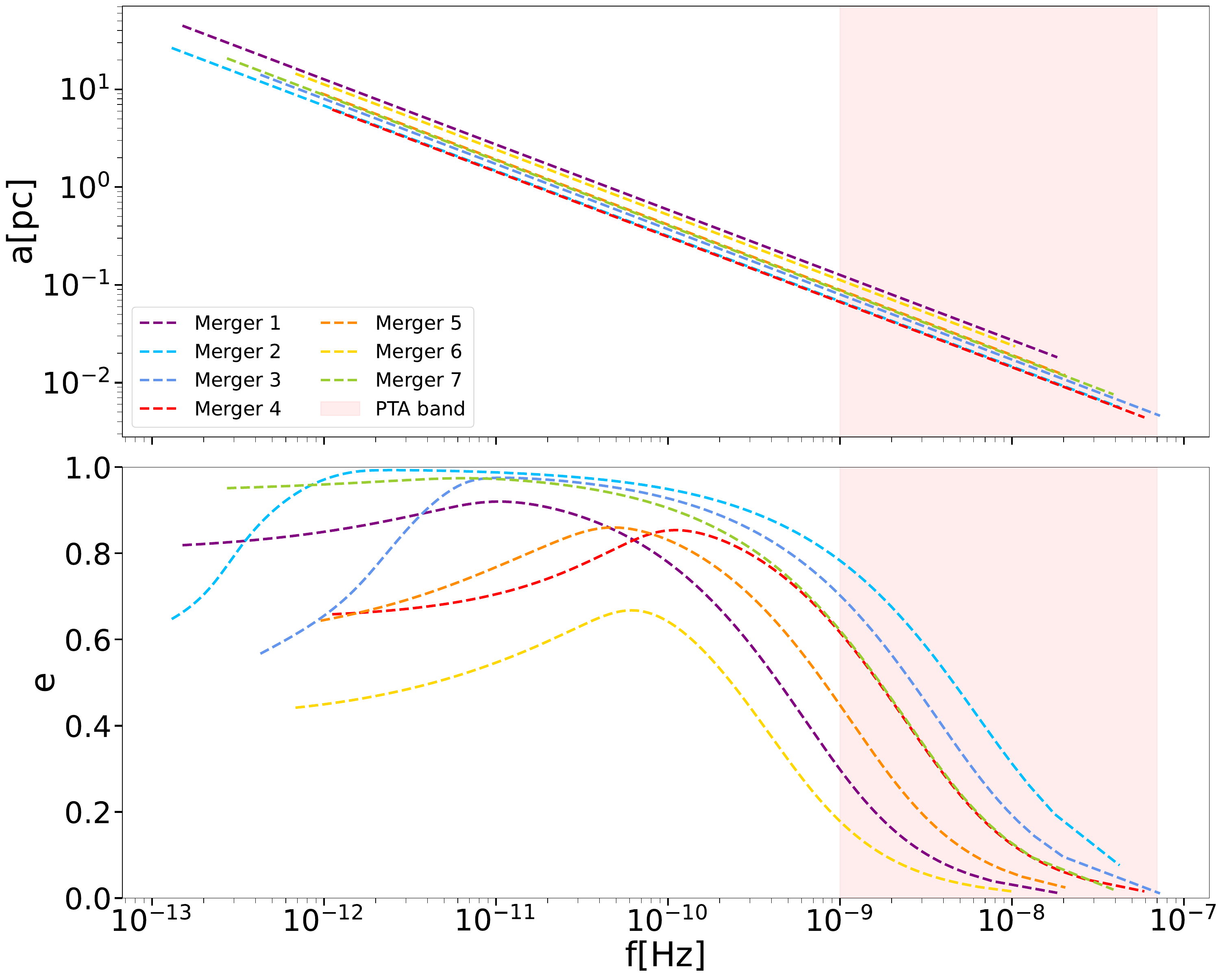}
    \caption{The semi-major axis (top panel) and eccentricity (bottom panel) of the BHBs formed in all of the selected mergers as a function of GW frequency. For visual clarity we plot only the semi-analytical evolution. We highlight the PTA frequency band with a shaded area.}
    \label{fig:freq}
\end{figure}

In order to highlight the relevance of our results for GWs detection by PTA, we computed the GW frequency of our binaries via $f_{\rm{GW}}=2f_{\rm{orb}}$, where $f_{\rm{orb}}$ is the the orbital frequency\footnote{We note that the value of $f_{\rm{GW}}$ thus computed is exact only if the binaries are circular. Moreover, we are not accounting for the redshift, though all of our mergers occur at low redshift, so that its effect is small and can safely be ignored for the sake of our discussion.}. In Fig. \ref{fig:freq} we plot the orbital parameters as functions of $f_{\rm{GW}}$ and highlight with a shaded area the frequency band relevant for PTA.

\begin{table*}
\centering

\setlength{\tabcolsep}{0.01\linewidth}
\begin{tabular}{|c | c c c c | c c c | c c c | c|} 
 
 \hline 

 ID & z & T$_{\rm{lb}}$ [Gyr] & $d_0$ [kpc] & $e_0$ & $t_b$ [Gyr] & $a_b$ [kpc] & $e_b$& $t_h$ [Gyr] & $a_h$ [kpc] & $e_h$ & $t_c$ [Gyr]\\ [1ex]

 \hline 
 1 & 0.197 & 2.479 & 82.81 & 0.553 & 0.62 & 0.12 & 0.56 & 0.72 & 0.04 & 0.83 & 5.1 \\[1ex]

 2 & 0.923 & 7.610 & 94.71 & 0.994 & 2.08 & 0.04 & 0.48 & 2.08& 0.04& 0.48 & 14.07 (3.71) \\[1ex]

 3 & 0.757 & 6.804 & 68.81 & 0.995 & 1.22 & 0.02 & 0.49 & 1.12& 0.03& 0.59 & 9.51(4.35) \\[1ex]

 4 & 0.676 & 6.350 & 46.33 & 0.995 & 0.46 & 0.04 & 0.20 &0.54&0.01&0.64 &  2.76 \\[1ex]

 5 & 0.440 & 4.741 & 78.28 & 0.976 & 0.53 & 0.12 & 0.38 &0.57&0.02&0.55 & 2.48 \\ [1ex]
 
 6 & 0.923 & 7.610 & 46.36 & 0.987 & 0.35 & 0.12 & 0.36 &0.38&0.03&0.31 & 3.7 \\ [1ex]

 7 & 0.169 & 2.168 & 60.12 & 0.886 & 0.47 & 0.12 & 0.87 &0.52&0.02&0.95 & 1.34 \\ [1ex]

 \hline
\end{tabular}
\caption{Parameters of each selected merger at critical times in the evolution. (i) Initial conditions of the {\tt Griffin} simulations: we report the initial redshift ($z$) of the merger drawn from TNG100-1 and its corresponding look-back time T$_{\rm{lb}}$, the initial distance of the BHs $d_0$ and the initial orbital eccentricity of the galactic merger $e_0$; (ii) time of BHB formation $t_{\rm{b}}$ and respective orbital parameters $a_b$ and $e_b$, (iii) time when the BHB reaches the hard-binary separation and orbital parameters $a_h$ and $e_h$. In the last column we report the predicted coalescence time. Times in brackets refer to semi-analytical models where the value of $K$ was increased to better match the eccentricity growth seen in the {\tt Griffin} simulations.}
\label{table:sum new complete}
\end{table*}

\section{Discussion} \label{discussion}

One of the main results of this work is that galactic mergers with mass ratios $q>0.1$ tend to happen on almost radial orbits, with $\sim 60\%$ of them having eccentricity $0.95\leq e < 1$. This is consistent with what has been reported  by \cite{2006A&A...445..403K} , who find that $\sim 40\%$ of the orbits have eccentricity values within the range of $1 \pm 0.1$. The small discrepancy can be attributed to a difference in setup and/or analysis: (i) they adopt a different cosmological simulation, performed within the {\tt GIF} project \citep{1999MNRAS.303..188K}, with different settings that affect the evolution and dynamics of structures; (ii) their study considers only DM halos, neglecting the stellar component when reconstructing the orbits; (iii) they calculate the orbital parameters at a time when the interaction between the halos is still weak and under the simplifying assumption of a two-body Keplerian system. On the other hand, we select the snapshot closest to the merger and calculate $M_{\rm{eff}}$ to define the Keplerian two-body system (see Sec. \ref{orbits method}). Our approach enables a reliable determination of the orbital parameters of the interacting galaxies, even for distances between galaxies under $100\kpc$. In turn, this implies more realistic initial conditions for the $N$-body simulations. 

We also find that binaries forming in almost radial mergers achieve a much lower eccentricity at binding. This may seem to contradict a previously reported correlation between the original eccentricity of the merger and the eccentricity at binary formation \citep{2022MNRAS.511.4753G}. However, the maximum initial orbital eccentricity considered in that study was $0.9$. If we limit our sample to mergers with $e_0\leq0.9$, we do find a similar trend where higher initial eccentricities lead to higher eccentricities at binary formation. We must note, however, that only mergers 1 and 7 in our sample satisfy the condition $e_0\leq0.9$, making this result subject to low number statistics.

\citet{2023MNRAS.526.2688R} find a significant scatter in eccentricity at binary formation when starting from mergers with $e_0=0.99$, with a weak dependence on the resolution of the simulation. They attribute this behaviour to the sensitivity of 
nearly radial trajectories to perturbations. Comparing our results is not straightforward as there are significant differences in the galaxy models and the initialisation of the $N$-body simulations. For instance, they consider idealised systems with mass ratio $q=1$ for both galaxies and black holes, they do not include DM halos in their simulations, and they model stellar bulges with a shallower inner slope compared to ours. In addition, they place the two galaxies at a very small initial separation of $3.72\kpc$. Nonetheless, our simulations have mass resolution similar to their medium-high mass resolution ($\mbh/m_{\star}\sim20000$) set and we observe some similarities in the peculiar behaviour of highly eccentric mergers which tend to result in less eccentric BHBs. Whether this effect can be reduced by further increasing the resolution is left to a follow up study. 
%AG here

Finally, we emphasise the fact that the BHBs orbital eccentricity can play a crucial role in explaining the observed flattening in the GWB spectrum.
A preponderance of highly eccentric mergers in the population of astrophysical binaries could be responsible for such flattening. The peculiar behaviour of moderate eccentricity in nearly radial mergers could be problematic in this regard. However, the physical mechanism responsible for lower eccentricity is not well understood, and will be investigated further. Nonetheless, the binaries in our sample show an eccentricity distribution ranging between $0.2$ and $0.8$ upon entry into the PTA frequency band (see Fig. \ref{fig:freq}), which could be sufficient to account for the flattening in the GWB spectrum. Moreover, current PTA data do not fully constrain the shape of the spectrum, and forthcoming results may refine our understanding of the current findings.

\section{Summary and Conclusions}\label{conclusions}
We have identified mergers of host galaxies to potential PTA sources in the cosmological simulation IllustrisTNG100-1. We aimed to determine the typical orbital eccentricity of these mergers, expecting to find a correlation between this and the eccentricity at the formation of BHBs. From the original sample we then selected a sub-sample of major mergers at low redshifts and studied the evolution of the BHB orbital parameters from formation down to separations of order a parsec by means of the FMM code {\tt Griffin}. We then employed a semi-analytical model to predict the evolution of the semi-major axis and eccentricity until coalescence.

Our main findings are as follows:
\begin{itemize}
    \item The majority of galactic mergers with $q>0.1$ at $z<2$  occur on nearly radial orbits, with $\sim 60\%$ of them having eccentricity $0.95\leq e < 1$. 
    \item The expected correlation between merger eccentricity and the eccentricity at binary formation holds up to a certain threshold: mergers with almost radial orbits ($e_0\sim0.99$) tend to form more circular BHBs.
    \item The eccentricity distribution upon entry in the PTA frequency band ranges from $0.2$ to $0.8$, which could still account for the observed flattening in the GWB at low frequencies.
    \item The majority of the simulated systems lead to final coalescence of the BHB within the Hubble time, thus resulting in potentially observable GW sources.
\end{itemize}

We plan to conduct a follow-up study to further investigate the physical processes behind the circularisation of BHBs formed after extremely eccentric encounters. Additionally, we expect that upcoming results from PTA collaborations will provide better constraints on the shape of the GWB spectrum, shedding light on the role of eccentricity in the evolution of supermassive black hole binaries.

\section*{Acknowledgements}

AS and EB acknowledge the financial support provided under the European Union's H2020 European Research Council (ERC) Consolidator Grant ``Binary Massive Black Hole Astrophysics'' (B Massive, Grant Agreement: 818691). AG would like to thank the Science and Technology Facilities Council (STFC) for support from grant ST/Y002385/1.
EB acknowledges support from the European Union's Horizon Europe programme under the Marie Skłodowska-Curie grant agreement No 101105915 (TESIFA).
The simulations were run on the Eureka2 HPC cluster at the University of Surrey.

%%%%%%%%%%%%%%%%%%%%%%%%%%%%%%%%%%%%%%%%%%%%%%%%%%
\section*{Data Availability}

The authors will share the data underlying this article upon reasonable request.



\bibliographystyle{mnras}
\bibliography{biblio.bib} % if your bibtex file is called example.bib

%%%%%%%%%%%%%%%%%%%%%%%%%%%%%%%%%%%%%%%%%%%%%%%%%%

%%%%%%%%%%%%%%%%% APPENDICES %%%%%%%%%%%%%%%%%%%%%

%\appendix


%If you want to present additional material which would interrupt the flow of the main paper, it can be placed in an Appendix which appears after the list of references.

%%%%%%%%%%%%%%%%%%%%%%%%%%%%%%%%%%%%%%%%%%%%%%%%%%

% Don't change these lines
\bsp	% typesetting comment
\label{lastpage}
\end{document}